\documentclass[a4paper]{article}

\usepackage{fullpage,url,amsmath,amsfonts,amssymb,tedmath,mathtools}

\newcommand{\M}{\mathcal}

\date{}
\title{Linear complementary dual,  maximum distance separable  codes\footnote{MSC 11T71, 94A24}}
\author{Ted Hurley\footnote{National University of Ireland
Galway. Ted.Hurley@NuiGalway.ie}} 
\begin{document}
\maketitle
\begin{abstract}
 Linear complementary dual (LCD)  maximum distance separable (MDS)   codes are constructed to given specifications. For given $n,r$, in which at least one of which is odd,  MDS LCD $(n,r)$ codes are constructed over any finite field whose characteristic does not divide $n$.  
 Series of LCD  MDS   codes are
constructed to required  rate and required error-correcting capability.  
For a given finite  field $GF(q)$ and given $n|(q-1)$,   LCD MDS codes of length $n$ and
 dimension $r$ are explicitly constructed over $GF(q)$ for all $r, (r<n)$
 when $n$ is odd and for all odd $r, (r<n)$ when $n$ is even.   
For given dimension and given error-correcting capability,  LCD MDS codes
 are constructed to these specifications and with smallest possible length. 
 Series of asymptotically good LCD MDS codes are explicitly constructed.
Explicit efficient encoding and decoding
 algorithms of complexity $\max\{O(\log n), O(t^2)\}$ exist for the constructed  LCD MDS codes where $t$ is the error-correcting capability of the code. 

  Linear complementary dual codes have importance in data storage, communications' systems and security.
 
\end{abstract}
\section{Introduction}

A Linear complementary dual,  LCD,  code is a linear code $\M{C}$ such that $\M{C}\cap \M{C}^\perp = 0$ where $\M{C}^\perp$ denotes the dual of $\M{C}$. 

LCD  codes have been studied extensively in
the literature. For background, history and general theory 
consult the nice  articles \cite{sihem3,sihem2,sihem4, sihem} by Carlet,
Mesnager, Tang, Qi and Pelikaan. LCD codes were originally introduced by Massey in \cite{massey,massey2}.
These codes have been studied for improving the security of information on sensitive devices against  {\em side-channel attacks} (SCA) and {\em fault non-invasive attacks}, see \cite{carlet}, and have  found use in   
{\em data storage} and {\em  communications' systems}. 

The necessary background on coding theory and field theory may be
found  in \cite{blahut} or \cite{mceliece}. The finite field of order $q$ is denoted here by $GF(q)$ and of necessity $q$ is a power of a prime.   

Here series of  LCD  MDS (maximum distance separable) codes are constructed to given requirements. For given odd $n$ and any $r<n$ and for given even $n$ and any odd $r<n$, MDS LCD $(n,r)$ codes are constructed over finite fields whose characteristic does not divide $n$.  
 For a given rate and given error-correcting capability, MDS LCD codes are
constructed to these specifications. 
For a given field $GF(q)$, and for each $n/(q-1)$, LCD MDS codes of length $n$ and dimension
 $r$ are explicitly constructed for any given $r$ ($r< n$) when $n$ is odd  and for any
 given odd $r$ ($< n$) when $n$ is even. The codes constructed  have  explicit efficient encoding and decoding algorithms.  

Carlet and Guilley \cite{carlet2} investigated the 
application of binary LCD codes against side-channel attacks (SCA) and
fault tolerant injection attacks (FIA). For the applications,  LCD codes with
specific dimension and specific error-correcting capability with length
as small as possible are required.  LCD MDS codes to required  dimension and required error-correcting capability to 
 smallest possible length for such a code are constructed here. 

Carlet and Guilley \cite{carlet2} also showed that
non-binary LCD codes in characteristic $2$ can be transformed into
binary LCD codes by expansion. Thus characteristic $2$ LCD codes are of particular importance. The results here include infinite series
of characteristic $2$ LCD MDS codes. 
For characteristic $2$ and given rate $R= \frac{r}{n}$ ($0< R < 1$) infinite series of LCD MDS codes  over finite fields of characteristic $2$ are constructed in which the limit of the ratio of the distance by the length approaches $(1-R)$. 

Infinite series of LCD MDS codes are constructed over fields of prime characteristic in which the ratio of distance by length is $(1-R)$ for given $R, 1 < R < 1$; in these cases the arithmetic is modular arithmetic with very efficient complexity of operations.  

The codes constructed are explicit and have explicit efficient encoding and decoding algorithms. These  
 follow from the algorithms in   \cite{hurley}. The complexity of encoding
 and decoding is
 $\max\{O(n\log n), O(t^2)\}$ where $n$ is the length and 
 $t=\floor{\frac{d-1}{2}}$, ($d$ the distance),  is the error-correcting  
 capability of the code.


Carlet, Mesneger, Tang and Qi \cite{sihem3} also construct some LCD MDS
codes.  Mesneger, Tang and Qi \cite{sihem} construct LCD codes using
algebraic geometry methods. The codes here  are different.  

An  {\em LCD Hermitian}  code is  defined as a code $\M{C}$ such that
$\M{C}\cap \M{C}^\perp_H = 0$ where $\M{C}^\perp_H$ denotes the dual of
$\M{C}$ relative to the   Hermitian product. Infinite series of
Hermitian LCD MDS 
codes may also be constructed over fields of the  form $GF(q^2)$ using
similar methods; the specifics  are omitted. 
\subsection{Layout} The general construction is given within Section \ref{construct}. The necessary background from \cite{hurley} is summarised  in Subsection \ref{background} and Subsection \ref{general} presents  the general constructions.  The complexity of encoding and decoding is discussed in Subsection  \ref{complexity}. Section \ref{char2}  constructs the codes over fields of characteristic $2$.  
Subsection \ref{similation} presents specific  simulations and samples
with which to demonstrate the main constructions. A method for
constructing LCD MDS codes to given rate and given error-correcting
capability is derived in Section \ref{gets} and samples are
given. Section \ref{carlet} gives the best length construction of LCD
MDS codes for given dimension and required error-capability. 
Section \ref{asym} deals with deriving asymptotic constructions:  for given rate $R$ series of MDS LCD codes are constructed with this rate and in which the limit of the distance by the length approaches $(1-R)$.  Subsection \ref{prime} derives such constructions over prime fields; Subsection \ref{cahr2} derives asymptotic such constructions over fields of particular characteristic $2$.  

\section{Construction}\label{construct} 

\subsection{Background material}\label{background} 
The codes constructed are based on methods of
\cite{hurley,hurley1}, using  the unit-derived schemes  of 
\cite{hur1} applied to Vandermonde and Fourier matrices. 

A primitive $n^{th}$ root of unity $\om$ in a field $\F$ is an element $\om$ satisfying $\om^n = 1_{\F}$ but $\om^i \neq 1_{\F}, 1\leq i < n$. The multiplicative identity of a field $\F$ will be denoted by $1$ rather than $1_{\F}$ when the field in question is clear. 
The field $GF(q)$  contains a primitive $(q-1)$ root of unity, \cite{blahut,mceliece}, and such a root  is referred to as a {\em primitive element in the field $GF(q)$}. In addition then the field $GF(q)$ contains primitive $n^{th}$ roots of unity for any $n/(q-1)$. 


An $(n,r)$ linear code is a linear code of length $n$ and dimension $r$;
the rate is $\frac{r}{n}$. An $(n,r,d)$ linear code is a code of length
$n$, dimension $r$ and (minimum) distance $d$. The code is an MDS code
provided $d=(n-r+1)$, which is the maximum distance an $(n,r)$ code can
attain. The error-capability of $(n,r,d)$ is $t=\floor{\frac{d-1}{2}}$ which is the
maximum number of errors the code can correct successfully. 


Let $\om$ be a primitive $n^{th}$ root of unity in a
field $\F$. The Fourier matrix $F_n$, relative to $\om$ and $\F$, is the $n\ti n$ matrix  \quad 
$ F_n= \begin{pmatrix}1 & 1 & 1& \ldots & 1 \\ 1 & \om & \om^2 & \ldots &
	 \om^{n-1} \\ 
1 & \om^2 & \om^4 & \ldots & \om^{2(n-1)} \\ \vdots & \vdots & \vdots &
    \ldots & \vdots \\ 1 & \om^{n-1} & \om^{2(n-1)} & \ldots &
    \om^{(n-1)(n-1)} \end{pmatrix}$. 


Then  $\begin{pmatrix}1 & 1 & 1& \ldots & 1 \\ 1 & \om & \om^2 & \ldots & 
	 \om^{n-1} \\ 
1 & \om^2 & \om^4 & \ldots & \om^{2(n-1)} \\ \vdots & \vdots & \vdots &
    \ldots & \vdots \\ 1 & \om^{n-1} & \om^{2(n-1)} & \ldots &
    \om^{(n-1)(n-1)} \end{pmatrix} \begin{pmatrix}1 & 1 & 1& \ldots & 1 \\ 1
				    & \om^{n-1} & \om^{2(n-1)} & \ldots
				    & \om^{(n-1)(n-1)}\\ 
1 & \om^{n-2} & \om^{2(n-2)} & \ldots &\om^{(n-1)(n-2)} \\ \vdots & \vdots & \vdots &
    \ldots & \vdots \\ 1 & \om & \om^{2} & \ldots &
    \om^{(n-1)} \end{pmatrix} = nI_n$. 

The second matrix on the left of the equation is denoted by $F_n^*$;
then $F_nF_n^*= nI_n$. This $F_n^*$ 
 can be obtained from $F_n$ by replacing $\om$ by $\om^{n-1}$ and is a Fourier matrix itself.
An $n^{th}$ root of unity can only
exist in a field whose characteristic does not
divide $n$ and then $\frac{1}{n}$ exists in that field to give $F_n^{-1} = \frac{1}{n}F_n^*$.

 Explicit series of asymptotically good LCD MDS codes are
constructed.  For given
rate $R$, $0<R<1$, infinite series of LCD MDS codes are constructed in
which the ratio of the distance by the length approaches $(1-R)$.

In \cite{hurley} rows of $F_n$ in specific sequence are used to 
generate MDS codes, with efficient encoding and decoding algorithms. 
The required results from \cite{hurley} are given as follows. 

\begin{theorem}{\cite{hurley}}\label{seq} Let $\mathcal{C}$ be a code generated by taking any $r$ rows of $F_n$ in arithmetic sequence with arithmetic difference $k$ satisfying $\gcd(n,k) = 1$. Then $\mathcal{C}$ is an MDS (maximum distance separable) $(n,r,n-r+1)$ code.
\end{theorem} 
In particular we have:
\begin{theorem}{\cite{hurley}}\label{seq1} Let $\mathcal{C}$ be a code generated by taking $r$ consecutive rows of $F_n$. Then $\mathcal{C}$ is an MDS $(n,r,n-r+1)$ code.
\end{theorem}

For  `rows in sequence' it is permitted that rows may {\em wrap around}
and then $e_k$ is taken to mean $e_{k \mod n}$. Thus for example Theorem \ref{seq1}
could be applied to a code generated by $\langle e_r, \ldots, e_{n-1},
e_0, e_1, \ldots, e_s\rangle$. 

There are similar theorems, see \cite{hurley}, involving the more general
Vandermonde matrices but these are not used here.  

Now $F_n^*$ denotes  the matrix with $F_nF_n^*= nI_{n\ti n}$. 
Denote the rows of $F_n$ in order by $\{e_0,e_1, \ldots, e_{n-1}\}$ and denote 
the columns of $F_n^*$ in order by $\{f_0,f_1, \ldots, f_{n-1}\}$. { Then it is important to note that}  
$f_{i} = e_{n-i}\T, e_i = f_{n-i}\T$ with the convention that suffices
are taken modulo $n$. 
Also note $e_if_{i}= n$ and $e_if_j = 0, i\neq j$.
  

 Thus 

$$\begin{pmatrix}e_0 \\ e_1 \\ \vdots \\ e_{n-1}\end{pmatrix} (f_0,
f_1, f_2, \ldots, f_{n-1})= \begin{pmatrix}e_0 \\ e_1 \\ \vdots \\ e_{n-1}\end{pmatrix} (e_0\T,
e_{n-1}\T, e_{n-2}\T, \ldots, e_1\T) = nI_n$$



\subsection{LCD MDS codes: General constructions}\label{general}
It is required to construct $(n,r)$ MDS LCD codes over finite fields. 

First construct a Fourier $n\ti n$ matrix $F_n$. There is some restriction,
depending on $n$, on the characteristic of the finite fields that may be used; this is made more precise in subsection \ref{type} below. 



Suppose a  dimension size $(2r+1)$ is required for the code. Let $\M{C} = \langle
e_0,e_1, \ldots, 
e_{r}, e_{n-r}, e_{n-r+1}, \ldots, e_{n-1} \rangle$. The generators of $\C$ can be
given in sequence as $\{e_{n-r}, e_{n-r+1}, \ldots, e_{n-1}, e_0, e_1,
\ldots, e_{r}\}$. Thus the code  $\M{C}$ satisfies the criteria of Theorem \ref{seq1}
which implies that $\M{C}$  is  an $(n,2r+1, n-2r)$ MDS code. Now $\M{C}*(f_{r+1}, f_{r+2},
\ldots, f_{n-r-1}) = 0_{(2r+1)\ti (n-2r+1)}$ (where $*$ denotes matrix multiplication).

As $(f_{r+1},f_{r+2}, \ldots ,
f_{n-r-1}) = (e_{n-r-1}\T, e_{n-r-2}\T, \ldots, e_{r+1}\T)$
this gives $\M{C}^\perp = \langle e_{r+1}, e_{r+2}, \ldots, e_{n-r-1}\rangle$ and so
$\M{C}\cap \M{C}^\perp = 0$. Thus $\M{C}$ is an $(n, 2r+1, n-2r))$
MDS LCD code.

When choosing the rows from the Fourier matrix $F_n$ for the generating
      matrix, it is noted that {\em if
      $e_i$ is chosen then also $e_{n-i}$ must
be chosen} when  an LCD code is required. For an MDS code to be
      constructed 
      the rows chosen must also be in arithmetic sequence with arithmetic
      difference $k$ satisfying $\gcd(k,n)=1$ in order to satisfy the
      conditions of Theorem  \ref{seq}.  

In general when choosing rows from the Fourier matrix to form the
generator matrix if the row $e_i$ is chosen then also the row $e_{n-i}$
must also be chosen when  an LCD code 
is required; to get an MDS code the rows should finish up in
sequence so that Theorem \ref{seq} may be applied. When $n$ is even it
is not possible by this method to get an even dimension $r$ and rows in
sequence to satisfy Theorem \ref{seq} as the arithmetic differences $k$ that can be obtained are also even and then $\gcd(n,k) \geq 2$. 

The above argument allows the construction of $(n,2r+1)$ LCD MDS codes whether $n$ is even or odd and for any (odd) $(2r+1)$. If $n$ is odd this also allows the construction of $(n,r)$ LCD MDS for any $r$ since if $\C$ is an LCD MDS code then so is $C^\perp$.

It is worthwhile to construct directly $(n,r)$ LCD MDS codes for odd $n$ and any $r$. If $r$ is odd then proceed as before. Suppose $r=2t$ is even and $n$ is odd. Define $\C= \langle e_1, e_3,\ldots,  e_{r-1}, e_{n-r+1}, e_{n-r+3}, \ldots, e_{n-1}\rangle$. Then $\C$ can be given in sequence $\langle e_{n-r+1}, e_{n-r+3}, \ldots, e_{n-1}, e_1,e_3, \ldots, e_{r-1}\rangle$ with arithmetic difference $2$; now $\gcd(n,2)=1$ and hence by Theorem \ref{seq} $\C$ is an MDS code. It is also seen to be an LCD code since if $e_i\in \C$ then also $e_{n-i} \in \C$. Thus $\C$ is and LCD MDS $(n,r)$ code.  

From a Fourier $n\ti n$ matrix other LCD MDS $(n,r)$ codes are obtained for given $r$ by varying the allowed arithmetic difference as per Theorem \ref{seq}; $r$ must be odd if $n$ is even.  Let $k < n, \gcd(k,n) = 1$ and suppose that $r$ is odd; when $r$ is even and $n$ is odd the constructions are similar.  Then 
define $\M{C} = \langle
e_0,e_k, \ldots, 
e_{rk}, e_{n-rk}, e_{n-rk+k}, \ldots, e_{n-k} \rangle$. Here the suffices $rk$ mean $r*k$ and all suffices are taken $\mod n$. The generators of $\C$ can be
given in arithmetic sequence as $\{e_{n-rk}, e_{n-rk+k}, \ldots, e_{n-k}, e_0, e_k,
\ldots, e_{rk}\}$ with arithmetic difference $k$ and $\gcd(k,n)=1$.  Thus the code  $\M{C}$ satisfies the criteria of Theorem \ref{seq}
which implies that $\M{C}$  is  an $(n,2r+1, n-2r)$ MDS code. As with the case $k=1$ above it is seen that $\M{C}$ is also an LCD code.

Thus $\M{C}$ as constructed is an $(n, 2r+1)$ LCD MDS code.

So for each $k$ satisfying $\gcd(k,n) = 1$ LCD MDS codes may be
constructed of the form $(n,2r+1)$. Thus $\phi(n)$ different such LCD MDS codes may be constructed from the Fourier $n\ti n$ matrix.


To construct an MDS LCD code of type $(n,r)$ by the methods here it is required that either $r$ or $n$ be odd. For a given {\em reduced} fraction $\frac{r}{n}$ either $r$ or $n$ is odd.

\subsection{Complexity}\label{complexity}
Efficient encoding and decoding algorithms exist for these codes by the
methods/algorithms  developed in \cite{hurley}. In general the complexity is  
$\max\{O(n\log n), O(t^2)\}$ where $n$ is the length and $t$ is the
error-correcting capability, that is,  $t= \floor{\frac{d-1}{2}}$ where
$d$ is the distance. 
See the algorithms in \cite{hurley} for details.


\subsection{Characteristic $2$}\label{char2}
Carlet and Guilley show \cite{carlet2} that
non-binary LCD codes in characteristic $2$ can be transformed into
binary LCD codes by expansion. Thus we look at characteristic $2$ in particular.  Let $GF(2^k)$ be a field of characteristic $2$. Then there exists a primitive $(2^k-1)$  root of unity $\om$ in $GF(2^k)$. Let $n=(2^k-1)$ and  $F_{n}$ be the Fourier $n \ti n$ formed using $\om$. Build LCD MDS codes of length $(2^k-1)$ and required rate as follows.

Let $(2r+1)$ be the dimension for the required rate. Then \\ 
let $\C = \langle e_0, e_1, \ldots, e_r, e_{n-r}, \ldots, e_{n-1}\rangle$ and  $\C$ is an LCD MDS $(n,2r+1)$ code by the general construction in section \ref{general} above.

Let $2r$ be the dimension for the required rate. Note that $n$ is
odd. \\  
Let $\C =\langle e_1, e_3, \ldots, e_{2r-1}, e_{n-2r+1}, e_{n-2r+3},
\ldots, e_{n-1}\rangle$. \\ This is in sequence $\{e_{n-2r+1},e_{n-2r+3},
\ldots, e_{n-1}, e_1, e_3, \ldots, e_{2r-1}\}$ with arithmetic
difference $2$ and $\gcd(n,2)=1$ and so satisfies the criteria of
Theorem \ref{seq}. Then $\C$ is an LCD MDS $(n,2r)$ code. 

There are as noted before other ways to construct the codes using $k$
with $\gcd(n,k)=1$ giving $\phi(n)$ different codes of required form. 
The size of the field must increase as the length increases. These are near {\em maximum length codes} for the field $GF(2^k)$.

Similarly for given $GF(2^k)$ LCD MDS codes of the form $(m,r)$ may be
constructed for any $m/(2^k-1)$ over this field. Note that $m$ is 
odd. 

\subsection{The  fields}\label{type} Suppose $n$ is given and it is required 
to find the the finite fields over which a Fourier $n\ti n$ matrix exists. 

\begin{proposition} There exists a finite field of characteristic $p$
 containing an $n^{th}$ root of unity for given $n$ if and only if
 $p\not\vert \, n$. 
\end{proposition}

This is well-known; a constructive proof is given as it  is needed later. 
 \begin{proof} Let
$p$ be a prime which does not divide $n$.  Hence  $p^{\phi(n)} \equiv 1
\mod n$ by Euler's theorem where $\phi$ denotes the Euler $\phi$ function. More specifically let $\be$ be the least
positive integer such that $p^{\be} \equiv 1
\mod n$. Consider $GF(p^\be)$. Let $\de$ be a primitive element in
$GF(p^\be)$. Then $\de$ has order $(p^\be-1)$ in $GF(p^\be)$ and $(p^\be -1) = s n$ for
some $s$. Thus $\om= \de^s$ has order $n$ in $GF(p^\be)$.

On the other hand if $p/n$ then $n=0$ in a field of characteristic $p$ and so no $n^{th}$ root of unity can exist in the field.

\end{proof}

This enables the construction of the smallest field of characteristic $p$, $p\not\vert \, n$, 
 with an $n^{th}$ root of unity. 
If $p$ is a prime not dividing $n$ find the least positive power (it exists by Euler's Theorem)  such that $p^\al \equiv 1 \mod n$ and then the field $GF(p^\al)$ contains an $n^{th}$ root of unity. The Fourier $n\ti n$ matrix over $GF(p^\al)$ may be constructed. 

\paragraph{Examples}\label{fields}
Suppose $n=52$. The prime divisors of $n$ are $2,13$ so take any other prime $p$ not $2,13$ and then there is a field of characteristic $p$ which contains a $52^{nd}$ root of unity. For example take $p=3$. Know $3^{\phi(52)} \equiv 1 \mod 52$ and $\phi(52) = 24$ but indeed $3^6 \equiv 1 \mod 52$. Thus the field $GF(3^6)$ contains a primitive $52^{nd}$ root of unity and the Fourier $52 \ti 52$ matrix exists in $GF(3^6)$. Also $5^4\equiv 1 \mod 52$, and so $GF(5^4)$ can be used.  Now $5^4 = 625 < 729 = 3^6$ so $GF(5^4)$ is a smaller field with  which to work.

Even better though is $GF(53)= \Z_{53}$ which is a prime field. This has an element of order $52$ from which the Fourier $52\ti 52$ matrix can be formed. Now $\om = (2 \mod 53)$ is an element of order $52$ in $GF(53)$. Work and codes with the resulting Fourier $52\ti 52$ matrix can then be done in modular arithmetic, within $\Z_{53}$, using powers of $(2 \mod 53)$. 

\subsection{LCD MDS sample cases.}\label{similation}
The best way to understand the general constructions is by looking at suitable relatively small samples and prototypes; but  in general there is no restriction on the length or on
the dimension for which MDS LCD codes can be constructed by the method. 
 
\paragraph{Length $7$.} 
Consider a Fourier $7\ti 7$ matrix $F_7$ over a
field $\F$. Suitable fields need to be  of characteristic not dividing $7$
and containing an element of order $7$. Thus   $GF(2^3), GF(3^6), GF(5^6), GF(11^2),
      GF(13^2), ... $ are suitable fields over which LCD MDS $(7,r)$ codes may be constructed by the methods. Now $GF(2^3)$ is
 of  characteristic $2$, which is good and may be required, and this also
is  the smallest field with a primitive $7^{th}$ root of unity.

Let $F_7$ be a Fourier matrix constructed from a primitive $7^{th}$ root of unity $\om$. 
Denote  the rows of  $F_7$ in order by $\{e_0,e_1,\ldots, e_6\}$ and
the columns of $F_7^*$ in order by 
$\{f_0,f_1,\ldots, f_6\}$. Then $e_if_j= 7\de_{ij}$ and $f_i= e_{7-i}\T$
with the convention that $e_7=e_0$. 

Construct $(7,3,5)$ and $(7,5,3)$ MDS LCD codes as follows. 
Let $\M{C} = \langle e_0, e_1, e_6 \rangle$. Notice that the rows of
$\M{C}$ are in sequence $\{e_6,e_0, e_1\}$ and so satisfy the conditions of
Theorem \ref{seq1}. Thus $\M{C}$ generates an  $(7,3,5)$ MDS code. 
Check that $\M{C}$ generates  an LCD code. Now $\M{C}*(f_2,f_3,f_4,f_5) = 0_{3\ti
4}$, $(f_2,f_3,f_4,f_5)= (e_5\T,e_4\T,e_3\T,e_2\T)$ (where $*$ is matrix
multiplication) and hence
$\M{C}^\perp = \langle e_2,e_3,e_4,e_5\rangle$. Thus $\M{C}\cap
\M{C}^\perp = 0 $ as required.
Hence $\M{C}$ is an $(7,3,5)$ MDS LCD code. 

Now let $\M{C} = \langle e_0, e_1,e_2, e_5 , e_6 \rangle$. Notice that the rows of
$\M{C}$ are in sequence $\{e_5, e_6,e_0,e_1,e_2\}$ and so satisfy the conditions of
Theorem \ref{seq1}. Thus $\M{C}$ is an $(7,5,3)$ MDS code. It is  now
necessary  to
show  that $\M{C}$ generates  an LCD code. Now  $\M{C}*(f_3,f_4) = 0_{5\ti
2}$, $(f_3,f_4)= (e_4\T, e_3\T)$ and hence
$\M{C}^\perp = \langle e_3,e_4\rangle$. Thus $\M{C}\cap
\M{C}^\perp  = 0 $ as required. 

Other LCD MDS codes may be constructed as follows. Note $\gcd(3,7)=1$. Thus Let $\C= \langle e_0, e_3,e_6, e_{7-6}, e_{7-3}\rangle = \langle e_0, e_3,e_6,e_1,e_4 \rangle = \langle e_1, e_4, e_0, e_3,e_6 \rangle$. Thus $\C$ satisfies the conditions of Theorem \ref{seq} and so is an MDS code. It is also an LCD code as is easily checked since if $e_i$ is included so is $e_{7-i}$. Thus this $\C$ is an LCD MDS $(7,5,3)$ code.

The method allows the construction of $\phi(7) = 6$ MDS LCD codes of
      form $(7, 5,3)$. 

Consider now constructing an LCD MDS code $(7,r)$ where $r$ is
      even. Consider $r=4$. Let $\C= \langle e_1, e_3, e_4,
      e_6\rangle$. This is given in sequence $\langle e_4,
      e_6,e_1,e_3\rangle$ with arithmetic difference $2$ and $\gcd(7,2)=
      1$. Hence by Theorem \ref{seq} $\C$ is an MDS code. It is also an
      LCD code and is therefore an MDS LCD $(7,4)$ code. 




Is there a prime field with an element of order $7$? Yes $GF(29)$
      has an element of order $7$. Then  $\om = (7 \mod 29)$ has order $7$ in $GF(29)=\Z_{29}$
      and may be used to form a Fourier $7\ti 7$ matrix in $GF(29)$. The calculations are then done in modular arithmetic -- in $\Z_{29}=GF(29)$.  
\paragraph{Length $11$.} 

Suppose a length $11$ LCD mds code
is required. Let $F_{11}$ be a Fourier $11\ti 11$ matrix over some field
$\F$. Appropriate fields include $GF(2^{10}), GF(3^5), GF(23)$ since an
element of order $11$ exists in these fields. Note that $GF(23)=\Z_{23}$
is a prime field and the arithmetic is modular arithmetic; for example  $2 \mod
23$ has order $11$ in $GF(23)$ and this (modular) element could be used
to generate the Fourier matrix. 

Now $(11,9,3),(11,7,5),(11,5,7),(11,3,9)$ MDS LCD codes 
can be generated codes with  $F_{11}$. 

Denote the rows in order of $F_{11}$ by $\{e_0, e_1, \ldots, e_{10}\}$ and the
columns in order of $F_{11}^*$ by $\{f_0,f_1,\ldots, f_{10}\}$. As noted $e_if_j=11
\de_{ij}, e_i\T=f_{11-i}$. 

Now let $\M{C}= \langle e_0,e_1, e_2,e_3, e_4, e_7,e_8,e_9,e_{10}
\rangle$. Then $\M{C}$ can be given in sequence
\\ $\{e_7,e_8,e_9,e_{10},e_0,e_1,e_2,e_3,e_4\}$ and hence satisfies the
criteria of Theorem \ref{seq} and thus is a $(11,9,3)$ MDS code. Then
$\M{C}*(f_5,f_6)= 0_{9\ti 2}$. Now $(f_5,f_6)= (e_6\T,e_5\T)$ and hence
$\M{C}^\perp = \langle e_5,e_6 \rangle$ and so $\M{C}\cap \M{C}^\perp =
0$. Thus $\M{C}$ is an $(11,9,3)$ MDS LCD code. 

The $(11,7,5), (11,5,7),(11,3,9)$ MDS LCD required codes are constructed
similarly.

The methods as noted  allows the construction of $\phi(11)=10$ LCD MDS codes of  each type.

By methods/algorithms of \cite{hurley} the codes   have efficient encoding and decoding
algorithms. 

Also $(11,r)$ MDS LCD codes for $r$ even are produced using $\langle e_1, e_3,
\ldots, e_8,e_{10}\rangle$ similar to the methods in the case of length $7$.

Exercises: Over what finite fields do $11 \ti 11$ Fourier matrices
exist? What is the smallest non-prime field over which an $11 \ti 11$
Fourier matrix exists?  
(See Section \ref{type}.) 
 


\paragraph{Relatively large sample with modular arithmetic.}

Consider $GF(257) =\Z_{257}$ and $257$ is prime. 
Construct the Fourier matrix $F_{256}$ with a primitive $256^{th}$ root of
      unity $\om$ in $GF(257)$. Since the order of $3 \mod 257$ is $256$
      then a choice for $\om$ is $ (3 \mod 257)$. Denote the rows of
      $F_{256}$ in order by $\{e_0, e_1, \ldots, e_{255}\}$.

Suppose a dimension $r=2t+1$ is required $t\geq 0$. Choose $\C = \langle e_0,
      e_1, \ldots, e_t, e_{256-t}, e_{256-t-1}, \ldots,
      e_{255}\rangle$. Then $\C$ generates an MDS LCD $(256,r)$
      code. The arithmetic is modular arithmetic, $\mod 257$, and is
      dealing  with powers
      of $(3 \mod 257)$. 

Note for example that $(5 \mod 257)$ or $(7 \mod 257)$ could also be
      used to generate  the Fourier $256\ti 256$ matrix. 


      The method allows the construction of $\phi(256)=128$ such LCD MDS $(256, 2r+1)$ codes. For larger primes the  number that can be construction is substantial and cryptographic methods can be devised. For example for the prime $p=2^{31}-1$ the Fourier $(p-1)\ti (p-1) $ Fourier matrix exists over $GF(p)$ and $\phi(p-2)=534600000$. 

\subsection{To required dimension and error-correcting
  capability}\label{carlet}

Let a dimension $k$ be given and it is required to construct LCD MDS
codes with this dimension and required error-correcting capability as
per \cite{carlet2}.  

Let $k$ be the required dimension and LCD MDS codes of the form
$(n,r,d)$ are required where $\floor{\frac{d-1}{2}}$ is the required
error-correcting capability. 

By the bound for codes it follows that $d\leq n-k+1$ from which $n\geq
k+d -1$. So $n=k+d-1$ is the smallest the length $n$ can be. 

Consider then $n=k+d-1$. Construct a Fourier $n\ti n$ matrix over a
finite field.  Such matrices may be constructed over a field of
characteristic not dividing $n$ and of large enough order so that the
field contains an element of order $n$, see section \ref{type}. 
If $k$ is odd or if $n$ is odd then by the general method of Section
\ref{general} LCD MDS codes may be constructed of the form $(n,k)$. The
distance of the code is $n-k+1 = d$ as required.
If both $n$ and $k$ are even then $n=k+d-1$ implies that $d$ is
odd. Replace $d$ by $d+1$ which is even but also $\floor{\frac{d-1}{2}}
= \floor{\frac{d}{2}}$ for odd $d$s require a length $n=k+d$ which is
odd. Now construct  LCD MDS $(n,k)$ codes by the methods of
Section{general} which has distance $n-k+1= d+1$ which has
error-correcting capability $\floor{\frac{d}{2}} =
\floor{\frac{d-1}{2}}$ as required. 

\subsubsection{Sample} Consider requiring a dimension $7$ which can
correct $3$ errors. Then a $(n,7,7)$ code requires $n\geq 13$. Let
$n=13$. Now construct a $13\ti 13$ Fourier matrix over a finite
field. Then by method of subsection \ref{general} construct the LCD MDS
code $(13,7,7)$. The dimension is $7$ as required and the
error-correcting capability is $\floor{\frac{7-1}{2}}=3$ as required. 

Over which fields may the Fourier $13\ti 13$ matrix be constructed? See
section \ref{type}. For
characteristic $2$ the field required is $GF(2^{12})$ as the order of $2
\mod 13$ is $12$. The field is large. Now the order of $3 \mod 13$ is $3$
and so the field $GF(3^3)$   could be used which is nice. 

By methods of \cite{carlet2} (Proposition 3) this $(13,7,7)$ LCD MDS code over $GF(3^3)$
could be expanded to a $(39,21,\geq 7)$ LCD code over $GF(3)$. However
LCD MDS $(39,21,19)$ codes may be obtained over other fields, for
example over $GF(2^12)$ or $GF(5^4)$, by the general methods of \ref{general}.  
 
Now $GF(53)$
contains an element of order $13$ so this prime field $GF(53)=\Z_{53}$
could be used with modular arithmetic. Then $\om = (10 \mod 53)$ has
order $13$ and so this (modular) element could be used to generate the
Fourier $13 \ti 13$ matrix over $GF(53)$. There are other (modular)
elements of  order
$13$ in $GF(53)$ from which the Fourier $13 \ti 13$ matrix can
be constructed.

\subsubsection{Samples 2} Suppose  a dimension $227$ is required which can correct $14$ errors. Thus
a $(n, 227, 29)$ code is required. Now $n\geq 227+29-1 = 255$. Let
$n=255$. Construct a Fourier $255 \ti 255$ matrix over a finite field
and then by general method of Subsection \ref{general} construct the LCD
MDS $(255,227)$ code over this field; this has distance $29$ and
error-correcting capability of $14$.

Over what fields may the Fourier $255\ti 255$ matrix be constructed? Now
the order of $2 \mod 255$ is $8$ so the field $GF(2^8)$ may be used and
has an element of order $255$ which in this case is a primitive element in
$GF(2^8)$.      

In the prime field $GF(257)$  the Fourier  $256\ti 256$ exists. By
method of Subsection \ref{general} construct the $(256,227)$ LCD MDS
code over $GF(257)$. This has distance $30$ and so has error-correcting capability of
$\floor{\frac{30-1}{2}}= 14$ also. Then $\om = (3 \mod 257)$ has order
$256$ in $GF(257)$ and may be used to generate the Fourier $256\ti 256$
matrix over $GF(257)=\Z_{257}$.

\subsection{Codes to given rate and error-correcting capability}\label{gets}
Suppose an LCD MDS code of rate $R= \frac{r}{n}$  is required which can correct $\geq
t$ errors. In reduced fraction form $n$ or $r$ is 
odd; both could be odd. Define $n_i= i*n, r_i=i*r$ for odd $i$. Let $i$
be the least such that $t\leq \floor {\frac{n_i-r_i}{2}}$. 
  Construct the Fourier $n_i\ti n_i$ matrix
over a suitable finite field and from this construct by the general
method the MDS LCD $(n_i,r_i)$ code. The rate is $R$ and the
error correcting capability is $\floor{(n_i-r_i)/2}\geq t$. Other more
direct ways may be applied in particular instances.

\subsubsection{Sample: Rate $\frac{5}{7}$ required.}
  Suppose a rate $R=\frac{5}{7}$ is required which can
correct $25$ errors. Thus an MDS LCD  code $(n,r,\geq 51)$ is required where
$\frac{r}{n} = \frac{5}{7}$. This gives  $51 = n-r +1 = n(1-R)+1$ thus requiring $ 50=
n(\frac{2}{7})\geq 50$. This requires $n= 175$ and $r=125$. So construct a $175
\ti 175$ Fourier matrix and define $\M{C}=\langle e_0, e_1, \ldots,
e_{62}, e_{113}, e_{114},\ldots, e_{174}\rangle$. The rows are in sequence and hence  $\M{C}$ is an $(175, 125, 51)$ MDS code by Theorem \ref{seq1}. That it is also an LCD code follows from the general method.

Exercise: Over which finite fields can a Fourier $175 \ti 175$ be constructed? What is the smallest finite field over which a Fourier $175\ti 175$ can be constructed? 
\subsubsection{Sample: Rate $\frac{7}{8}$ required.}
Suppose a rate $R=\frac{7}{8}$ is required which can correct correct $25$ errors. Require a code $(n,r, \geq 51)$ such that $\frac{r}{n} = \frac{7}{8}$. Then require $n-r+1 = n(1-R) +1 = n(1-\frac{7}{8})+ 1 = n(\frac{1}{8}) +1\geq 51$.
Thus $n=400, r=350$.  Hence  a $(400, 350, 51)$ code is required -- or better.
Now the general method requires the dimension to be odd. Thus require a $(400,349, 52)$ code which has slightly less rate than $R$, which is okay for a system which can transmit at rate $R$, and this can correct $25$ errors.

Now $401$ is prime so the field $GF(401)$ has an element of order $400$ from which the Fourier $400\ti 400$ matrix may be defined over the prime field $GF(401)$. A suitable element of order $400$ in $GF(401)$ is $(3 \mod 401)$. 

A code with characteristic $2$ may be required so then consider constructing a $(401, 351, 51)$ code which has rate  slightly greater than $\frac{7}{8}$ 
or a $(401, 349, 53)$ code which has rate  slightly less than $\frac{7}{8}$.
Here though the order of $2 \mod 401$ is $200$ so would require a field of order $2^{200}$ to get an element of order $401$. However the order of $2 \mod 399$ is $18$ requiring a field $GF(2^{18})$. Form the Fourier $399 \ti 399 $ matrix in
$GF(2^{18})$ and use the general method; then deduce  an MDS LCD $(399,
349, 51)$.  This has rate  $0.87468..$ and can correct $25$ errors.  

Using the general method as described requires a $(8i,7i)$ MDS LCD code
such that $8i-7i\geq 50$. This requires $i\geq 50$ but requiring odd $i$
gives $i=51$ as the least such. Thus construct a $(408,357,51)$ LCD MDS
code of rate $\frac{7}{8}$ and can correct $25$ errors. Now $409$ is
prime so the field $GF(409)$ may be used in which there exists a $408$
root of unity. A solution is $\om = (21 \mod 409)$ and  the order of
this $\om$ is $408$ in $GF(409)$ and may be used to construct the
$408\ti 408$ Fourier matrix over $GF(409)$ from which the MDS LCD $(408,
357)$ code may be derived. 

\subsubsection{Rate $R= \frac{4}{5}$ in prescribed field type.}

Suppose a rate of  $R= \frac{4}{5}$  is required in a field of
characteristic $2$ for an MDS LCD
code which can correct at least $25$ errors. An $(n,r, n-r+1)$ MDS LCD  code
is required and the $n$ must be odd in order to find characteristic $2$ field with an element of order $n$. Then $d= n-r+1
\geq 51$ in order to correct $25$ errors. Thus $n(1-\frac{4}{5}) \geq 50$
which means $n\geq 250$. Take $n=255$ and then the field $GF(2^8)$  has
an element of order $255$. Now $255*\frac{4}{5}= 204$. But we need the
dimension to be odd in order for the general method to work to get an LCD code. . Take
$r=203$ and then the $(255,203, 53)$ code has rate $\frac{203}{255}=
0.7960..$ which is slightly less than $\frac{4}{5}$, which is fine as
the given system can transmit at rate $\frac{4}{5}$. 

Thus construct the Fourier $255\ti 255$ code over $GF(2^8)$ using the
primitive $255$ root of unity. Then construct the MDS LCD $(255,203)$
code by the general method above:

$\C= \langle e_0, e_1 ,\ldots, e_{101}, e_{154}, e_{155}, \ldots,
e_{254}\rangle$. Then $\C$ is an MDS LCD $(255,203)$ code over $GF(2^8)$.

Suppose a rate of at least $R= \frac{4}{5}$ is required in a prime field 
for an MDS LCD
code which correct at least $25$ errors. An $(n,r, n-r+1)$ MDS LCD  code
is required.  Then $d= n-r+1
\geq 51$ in order to correct $25$ errors. Thus $n(1-\frac{4}{5}) \geq 50$
which means $n\geq 250$. The next prime $\geq n+1$ is $257$. This has an
element of order $256$. Now $\frac{4}{5}*256$ is required for the
dimension which must be odd for the construction in section \ref{general} and hence require $r=205$. 
Then the code $(256,205,51)$ created over $GF(257)$ has rate $\frac{205}{256}$ which is $0.8007..$ only slightly greater than the required  rate.

\section{Asymptotics}\label{asym} Construct an infinite  series of LCD MDS codes
$(n_i,r_i)$ codes where the rate $\frac{r_i}{n_i} = R = \frac{r}{n}$ (fixed) 
with  $0 < R < 1$. Then for such a series it would
follow that $\lim_{i\rightarrow \infty} = 1-R$. 

We can assume that either $r$ or $n$ in the fraction $R= \frac{r}{n}$ is odd.

From a Fourier $n\ti n$ matrix construct an MDS LCD $(n,r)$ code; as
either $n$ or $r$ is odd this can be done by the general method of
section \ref{general}. Let
$n_1=n, r_1=r$ and define $n_i= i*n$ and $r_i=i*r$ for odd $i$ that is
for $i=
1,3,5,\ldots$. Then either $n_i$ or $r_i$ is odd. Construct the $n_i \ti
n_i$ Fourier matrix over some finite field.  By the general method,
from this  Fourier matrix construct an $(n_i,r_i) $ LCD MDS code. The rate
of the code is $\frac{r_i}{n_i} =  R$. The distance of the code is $
(n_i-r_i+1)$ and the ratio of the distance to the length is
$\frac{n_i-r_i+1}{n_i} = 1-R+ \frac{1}{n_i}$. As $i\rightarrow \infty$
this ratio of the distance to the length approaches $(1-R) $.     

Note that $0< R < 1$ if and only if $0< (1-R) < 1$ so could obtain an
infinite series of LCD MDS codes in which the ratio of the distance to
the length is required to  approach  a specific $R$ with $0< R < 1$.  

The explicit codes produced have efficient encoding and decoding algorithms by the algorithmic methods in  \cite{hurley}.
\subsection{Modular arithmetic; prime fields}\label{prime} Let $p$ be a prime and consider $GF(p)$. Then The Fourier matrix of size $(p-1)\ti (p-1)$ may be formed over $GF(p)$. Suppose a rate $R$ LCD MDS code is required over $GF(p)$.  Let $r = \floor{(p-1)*R+1}$ or $r=\floor{(p-1)*R}$ so that $r$ turns out to be odd. Construct the $(p-1, r) $ LCD MDS code by the general method.  Then the code has rate close to $R$ and the distance is $(p-1-r+1)=p-r$.

  Consider an infinite set of prime $p_1, p_2, \ldots$. Then construct for each $p_i$ an LCD MDS code $\C_i$ close to the given rate $R$ over the field $GF(p_i)$. Then the limit of these codes as $i\rightarrow \infty$ is $(1-R)$. 

  For example take all the primes of the form $4n+1$, namely $5,13, 17, 29, \ldots$ and require a rate of $R=\frac{3}{4}$.  In $GF(4n+1)= \Z_{4n+1}$ construct the Fourier $4n\ti 4n$ matrix using a primitive $(4n)^{th}$ root of unity.  Let $r = 3n$ if $n$ is odd and $r=3n - 1$ when $n$ is even. By the general method construct LCD MDS $(4n,r)$ codes. These are  LCD MDS codes $(4,3,2), (12,9,4), (16, 11, 6), (28,21,8),   (36, 27, 10), (40,29,12), (52,39,14) \ldots $  over the prime fields $GF(5), GF(13), GF(17), GF(29),GF(37), GF(41), GF(53), \ldots$ respectively. If the exact rate $\frac{3}{4}$ is required then only include primes of the form $4n+1$ where $n$ is odd.  The limit of the ratio of the distance to the length of this series of  codes  is $(1-R) = \frac{1}{4}$. 

\subsection{Characteristic $2$}\label{cahr2}
Suppose characteristic $2$ is required. Then it is required that the $n_i$ are
odd so  that a Fourier $n_i\ti n_i$ may be constructed over a field
of characteristic $2$. Thus for $R=\frac{r}{n}$ it is required that $n$ be odd and then proceed as in section \ref{asym}.

Here is  an example with characteristic $2$. Let $R=\frac{5}{7}$. An infinite series of LCD MDS
with this ratio is required where the ratio of the distance to the
length approaches $(1-R)$. Let $n_i = i*7$ and $r_i=i*5$ for $i=1,3,5,
\ldots$.  Now $n_i$ is odd so there exists a
Fourier $n_i \ti n_i $ matrix in a field of characteristic
$2$. Construct the LCD MDS code $(n_i,r_i)$ over this field by the
general method of section \ref{general}. The code
has rate $\frac{5}{7}$ for all $i$. As $i \rightarrow \infty$ the ratio of the
distance to the length approaches $(1 - \frac{5}{7}) = \frac{2}{7}$. The
codes are the LCD MDS codes $\{(7,5), (21,15), (35, 25), \ldots \}$
over different fields of characteristic $2$. For example the $(7,5)$ code could be
over $GF(2^3)$, the $(21, 15)$ code  could be over $GF(2^6)$, and the
$(35,25)$ code could be over $GF(2^{12})$ and so on. 

\end{document}